\documentclass[final,5p,times,twocolumn]{elsarticle}
\usepackage{amsfonts}
\usepackage{amsmath}
\usepackage{amsthm}
\usepackage{amssymb}
\usepackage{graphicx}
\usepackage{dcolumn}
\usepackage{bm}
\usepackage{bbding}
\usepackage{mathrsfs}
\usepackage{subfigure}
\usepackage{graphicx}
\usepackage{color}
\usepackage{xcolor}
\usepackage{epstopdf}
\usepackage{epsfig}
\usepackage[colorlinks,citecolor=blue,linkcolor=blue,hyperindex]{hyperref}
\DeclareMathOperator{\Trr}{Tr}
\biboptions{numbers,sort&compress}

\begin{document}

\begin{frontmatter}
\title{Flexible and experimentally feasible shortcut to quantum Zeno dynamic passage}
\author{Wenlin Li$^1$}
\author{Fengyang Zhang$^{1,2}$}
\author{Yunfeng Jiang$^3$}
\author{Chong Li\corref{cor1}$^1$}
\cortext[cor1]{Corresponding author. lichong@dlut.edu.cn}
\author{Heshan Song\corref{cor2}$^1$}
\cortext[cor2]{Corresponding author. hssong@dlut.edu.cn}
\address{$^{1}$ School of Physics and Optoelectronic Technology, Dalian University of
Technology, Dalian 116024, China}
\address{$^{2}$ School of Physics and Materials Engineering, Dalian Nationalities University, Dalian 116600, China}
\address{$^{3}$ Materials Science and Engineering, University of California, San Diego, 9500 Gilman Drive, La Jolla, CA 92093-0418, USA}
\begin{abstract}
We propose and discuss a theoretical scheme to speed up Zeno dynamic passage by an external acceleration Hamiltonian. This scheme is a flexible and experimentally feasible acceleration because the acceleration Hamiltonian does not adhere rigidly to an invariant relationship, whereas it can be a more general form $\sum u_{j}(t)H_{cj}$. Here $H_{cj}$ can be arbitrarily selected without any limitation, and therefore one can always construct an acceleration Hamiltonian by only using realizable $H_{cj}$.
Applying our scheme, we finally design an experimentally feasible Hamiltonian as an example to speed up an entanglement preparation passage. 
\end{abstract}
\end{frontmatter}

\section{Introduction}
In order to achieve some appropriate approximate conditions and to simplify physics system, the evolution speed has to be sacrificed sometimes, which was a common practice in quantum information processing (QIP). For a success QIP, however, one necessary prerequisite is that the evolution time should be short enough to avoid the influence of decoherence \cite{1,2}. An ideal solution for reconciling this contradiction is to append an external Hamiltonian in the system in order to ensure that the evolutions are similar to the results adopted approximation conditions. This basic principle of approximation acceleration had been applied to speed up adiabatic passages successfully in recent years \cite{3,4,5,6,7,8,9,10}, but the discussions about the accelerations on other approximations are still rare in both theoretical and experimental areas. In addition, another common defect of existing acceleration schemes is that almost all acceleration Hamiltonians are designed based on a fixed expression ($H_1(t)=i\hbar\sum_{n}\vert\partial_t\lambda_n\rangle\langle\lambda_n\vert$). There remain some difficulties in achieving those schemes in experiments because the corresponding acceleration Hamiltonian may consist of some non--physical interactions. For examples, Chen's scheme \cite{3} needed a transition between two ground states of a $\Lambda$-type atom and Lu's scheme \cite{6} required a swap-gate like term $\vert gf\rangle_{12}\langle fg\vert+H.c.$ in his acceleration Hamiltonian. Detuning driving fields may realize some of those non--physical interactions to some extent \cite{6,11,12}. But obtaining an effective interaction is bound to introduce other approximate conditions. Therefore, it is still an open question to design an acceleration Hamiltonian by only using reasonable interactions.

In this letter, we try to improve above two defects, i.e., (a): The acceleration scheme is extended  into other common approximations; (b): A general scheme is proposed (we call it ``flexible scheme'') so that the acceleration Hamiltonian, for a certain system, can be always divided to allowed interactions. We believe that such a scheme is universal and feasible in experiments. 

In recent years, quantum Zeno dynamic \cite{13,14} was also an approximation used widely to simplify Hamiltonian in entanglement preparation or quantum gate realization in a long evolution time ($gt\sim 10^2$) \cite{19,20,15,16,17,18,r1}. Unlike adiabatic approximation, Zeno approximation acceleration only requires the system to evolve into a specific subspace instead of a specific state. In other words, Zeno approximation acceleration corresponds to a more relaxed restriction and it is more suitable for a flexible designs. Thus, in this letter, we discuss how to speed up a Zeno dynamic process in detail, and present an example of the entanglement preparation to explain the fixed scheme and flexible scheme more intuitively. We demonstrate that the evolution time takes on an obvious reduction after the acceleration, and boundaries of decay rates are also relaxed. And above all, the generators of the flexible acceleration Hamiltonian in our example are exactly the ones of system Hamiltonian, which provides a promising platform for advancing the maneuverability of QIP.

Before the in-depth discussion, we firstly give a brief introduction about the Zeno dynamic and quantum Lyapunov control. Suppose a dynamical evolution of the whole system is governed by the Hamiltonian $H=H_0+H_I=H_0+KH_m$, where $H_0$ is the subsystem Hamiltonian to be investigated and $H_I=KH_m$ is an additional interaction Hamiltonian to perform the continuous coupling with the constant $K$. Under the strong coupling condition $K\rightarrow\infty$, the subsystem  investigated is dominated by the time--evolution operator ($\hbar=1$) \cite{13,14}:
\begin{equation}
U_0(t)=\lim_{K\rightarrow\infty}\exp(iKH_mt)U(t).
\label{eq:subtime}
\end{equation}
On the other hand, the time--evolution operator of this subsystem can also be expressed as: $U_0(t)=\exp(-it\sum_nP_nH_0P_n)$, where $P_n$ is the eigenprojection of the $H_m$ with the eigenvalue $\lambda_n$. The time--evolution operator of the whole system can then be simplified as:
\begin{equation}
\begin{split}
U(t)&\sim\exp(-iKH_mt)U_0(t)\\
&=\exp\left[-it\sum_{n}(K\lambda_nP_n+P_nH_0P_n)\right],
\label{eq:time}
\end{split}
\end{equation}
and we can obtain an effective Hamiltonian in the following form: $H_{eff}=\sum_{n}(K\lambda_nP_n+P_nH_0P_n)$. Because the Zeno condition requires a weaker $H_0$ compared with $H_m$, the evolution time will be quite long in this case.

{For a quantum system, the aim of quantum control is to make the system evolve to a specified target quantum state (or a target subspace) by designing appropriate time-varying control fields. The core idea of quantum Lyapunov control is to design an auxiliary function $V$ involving both quantum state and control field. $V$ can be regarded as a Lyapunov function if $V\geqslant 0$ and the system converges to the target state given by its saddle point $V=0$ \cite{21,21ad1,21ad2}. The Lyapunov control theory has demonstrated that the system can be controlled into the target state (subspace) only if the control fields are designed to meet $\dot{V}\leqslant 0$, i.e., the Lyapunov function is a monotonically nonincreasing function in the time domain corresponding to whole evolution process, and it tends to its minimum finally with the help of control fields.} 

\section{General formalism of shortcut scheme}
\label{subsection:General formalism}

\subsection{``Rough'' acceleration}
Similarly to adiabatic approximation acceleration, a simple idea of Zeno acceleration is to compensate the terms neglected in the approximate processes, in other words, those terms reappear in system Hamiltonian, that is, 
\begin{equation}
H_{R}=UH_{eff}U^{\dagger}-H_0-H_I,
\label{eq:Rough}
\end{equation}
consequently, the total Hamiltonian $H_t=H_0+H_I+H_R$ will be precisely equal to $H_{eff}$ after a diagonalization but without any approximation. Therefore some restrictions of key parameters are no longer necessary, which provides a prerequisite for the process of acceleration.

We call this scheme as ``rough'' acceleration  because the acceleration Hamiltonian based on this idea is irreplaceable, and it may be meaningless if some non–physical interactions exist in its expression. Oppositely, in the next subsection, we will introduce a more flexible scheme in which the acceleration Hamiltonian can be changed and replaced almost without limitation.
\subsection{Flexible acceleration}
Different rom the ``Rough'' acceleration, our aim is to realize an acceleration process with a more optional Hamiltonian. Consequently, the difficulties of the corresponding experiments will be significantly reduced because this acceleration Hamiltonian is not limited in the form of Eq. (\ref{eq:Rough}). For a general discussion, the external acceleration Hamiltonian can be written as $H_F=\sum_{j=0}^{n}u_{j}(t)H_{cj}$, where $H_{cj}$ can be chosen flexibly by the designers and $u_{j}(t)$ represent the corresponding control fields. A closed Zeno quantum system with this external acceleration Hamiltonian can be described by the following Liouville equation
\begin{equation}
\dfrac{d\rho}{dt}=\dot{\rho}=-i\left[\left(H_{0}+H_{I}+\sum_{j=0}^{n}u_{j}(t)H_{cj}\right),\rho\right].
\label{eq:Flexible}
\end{equation}
In order to obtain a suitable acceleration Hamiltonian, we define an auxiliary function $V(t)$ as:
\begin{equation}
V(t)=\Trr[H^{2}_{I}\rho(t)].
\label{eq:Auxiliary function}
\end{equation}
{We mark the $i$th eigenvalue and eigenvector of $H_I$ as $E_i$ and $\vert E_i\rangle$, respectively. Then Eq. \eqref{eq:Auxiliary function} becomes $V(t)=\sum_iP_i E^2_i$, i.e., the weighted sum of the eigenvalue squares. Here $P_i=\langle E_i\vert\rho(t)\vert E_i\rangle$ are the populations of $\rho(t)$ on the $i$th eigenvector. Under this definition, $V\geqslant 0$ is obvious. Such as the above-mentioned, Zeno approximation requires system to evolve in a Zeno subspace with degenerate eigenvalues. According to whether the corresponding eigenvectors are in the target space or not, we divide the populations into two parts,
\begin{equation}
\begin{split}
(\underbrace{P_1^T,P_2^T,...,P_m^T}_{target\,space},\underbrace{P_{m+1}^N,P_{m+2}^N,...,P_n^N}_{non\,target\,space}).
\label{eq:order}
\end{split}
\end{equation}
If the goal subspace selected corresponds to $E_i=0$ ($i\leqslant m$), $V(t)=0+\sum_{i>m}P^N_i E^2_i$ will be obtained. When the quantum state $\rho(t)$ is completely in the target space, $P^N_i=0$ are satisfied for all $i>m$. In other words, the system will be trapped in the goal subspace if and only if $V=0$, and correspondingly, the physical meaning of $V$ can be regarded as a violation measure of the Zeno subspace limitation.}

Substituting Eq. (\ref{eq:Auxiliary function}) into Eq. (\ref{eq:Flexible}), we can obtain the derivative of $V$ as follow:
\begin{equation}
\begin{split}
\dot{V}&=\Trr(H^{2}_{I}\dot{\rho})=\Trr\left(-iH^{2}_{I}\left[(H_0+H_I+\sum_{j=0}^{n}u_{j}(t)H_{cj}),\rho\right]\right)\\&=\Trr(-i\rho[H^{2}_I,H_0])+\sum_{j=0}^{n}u_{j}(t)\Trr(-i\rho[H_I^2,H_{cj}]).
\label{eq:dot function1}
\end{split}
\end{equation}
If the control fields are set as:
\begin{equation}
\left\lbrace
\begin{array}{ll}
u_0=-\dfrac{\Trr(-i\rho[H^{2}_I,H_0])}{\Trr(-i\rho[H_I^2,H_{c0}])}\\
\\
u_{j\neq 0}=-k_{j}\Trr(-i\rho[H_I^2,H_{cj}]),
\label{eq:CON function1}
\end{array}
\right.
\end{equation}
{and the second term in Eq. \eqref{eq:dot function1} can be expanded as 
\begin{equation}
\begin{split}
&\sum_{j=0}^{n}u_{j}(t)\Trr(-i\rho[H_I^2,H_{cj}])\\=&u_{0}(t)\Trr(-i\rho[H_I^2,H_{c0}])+\sum_{j=1}^{n}u_{j}(t)\Trr(-i\rho[H_I^2,H_{cj}])\\=&-\Trr(-i\rho[H^{2}_I,H_0])+\sum_{j=1}^{n}u_{j}(t)\Trr(-i\rho[H_I^2,H_{cj}])
\label{eq:adfunction1}
\end{split}
\end{equation}
by substituting $u_0$ in Eq. \eqref{eq:CON function1} into Eq. \eqref{eq:dot function1}.} Then Eq. (\ref{eq:dot function1}) can be simplified as $\dot{V}=-\sum_{j=1}^{n}k_{j}\Trr(-i\rho[H_I^2,H_{cj}])^2$ and it will always be negative if $k_{j}\geqslant 0$. By virtue of Eq. (\ref{eq:CON function1}), $V\geqslant 0$ and $\dot{V}\leqslant 0$ are satisfied simultaneously. In this case, $V$ is a so-called Lyapunov function, which can ensure that the system will evolve to the state corresponding to its own saddle point ($V=0$) \cite{21}. In other words, the acceleration Hamiltonian $H_F$ will control system into the goal subspace even if the Zeno condition is no longer satisfied. 

\section{Example in entanglement preparation}
\begin{figure}[]
\centering
\begin{minipage}[b]{0.5\textwidth}
\label{fig:subfig:a}   
\includegraphics[width=3.3in]{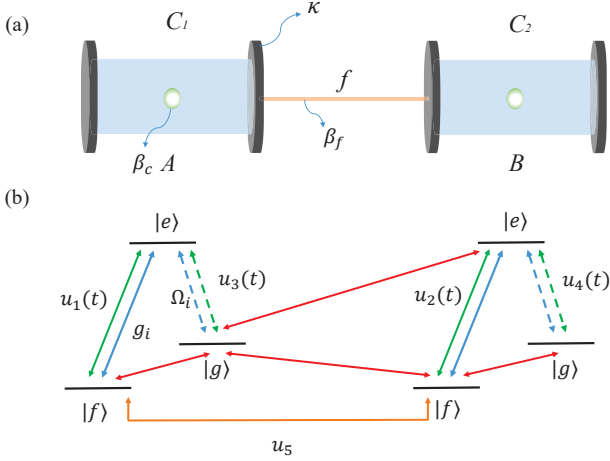}  
\end{minipage}
\caption{(a): Two $\Lambda$--type atoms ($A$, $B$) coupled with radiation fields in two resonant cavities ($C_1$, $C_2$). The fiber $f$ links two cavities. Here we encode as follow: $\vert\psi\rangle_i$ means that the $i$th subsystem is in the state $\vert\psi\rangle$, where $i\in\{A,B,C_1,C_2,f\}$. (b): The field--atom interaction $\vert f\rangle\leftrightarrow\vert e\rangle$ is provided by quantum fields, correspondingly, $\vert g\rangle\leftrightarrow\vert e\rangle$ is provided by classical fields. 
\label{fig:fig1}}
\end{figure}

Entanglement preparation is one of the most important issues in the field of QIP \cite{22,23}. In this section, we introduce and analyze a general entanglement preparation scheme based on Zeno dynamic to show the necessity of the shortcut and to explain our acceleration scheme in more detail. In the frame of cavity quantum electrodynamic (QED) system, the sketch of this entanglement preparation scheme is shown in Fig. \ref{fig:fig1}.
In the interaction picture, the Hamiltonian of the whole system can be described in the following form: $H=H_{laser}+H_I$, where
\begin{equation}
\begin{split}
&H_{laser}=\Omega_1\vert e\rangle_{A}\langle g\vert+\Omega_2\vert e\rangle_{B}\langle g\vert+H.c.\\
&H_{I}=g_1a_1\vert e\rangle_{A}\langle f\vert+g_2a_2\vert e\rangle_{B}\langle f\vert+\lambda[b^{\dagger}(a_1+a_2)]+H.c..
\label{eq:Hamilton}
\end{split}
\end{equation}
In above expressions, $a_1$($a_1^\dagger$) and $a_2$($a_2^\dagger$) are the annihilation(creation) operators of the cavity fields $C_1$ and $C_2$, respectively. $b$($b^\dagger$) is the annihilation(creation) operator of the fiber. $g_{1,2}$ and $\Omega_{1,2}$ are the coupling intensities respectively corresponding to the field--atom interactions $\vert f\rangle\leftrightarrow\vert e\rangle$ and $\vert g\rangle\leftrightarrow\vert e\rangle$, $\lambda$ is the coupling intensity of the fiber. Here we set $g_1=g_2$ for convenience. {In this model, $u_{j}$ between energy levels corresponding to green lines in Fig. \ref{fig:fig1} can be designed as time-dependent because the intensity of classical and quantum fields can be adjusted. $u_j$ denotes the fiber coupling intensities and it should be set as a fixed value. $u_j$ corresponding to all other transitions (e.g., red lines in Fig. \ref{fig:fig1}) should be zero because those transitions are not allowed physically.}
\begin{figure}[]
\centering
\begin{minipage}[b]{0.5\textwidth}
\label{fig:subfig:a}   
\includegraphics[width=3.5in]{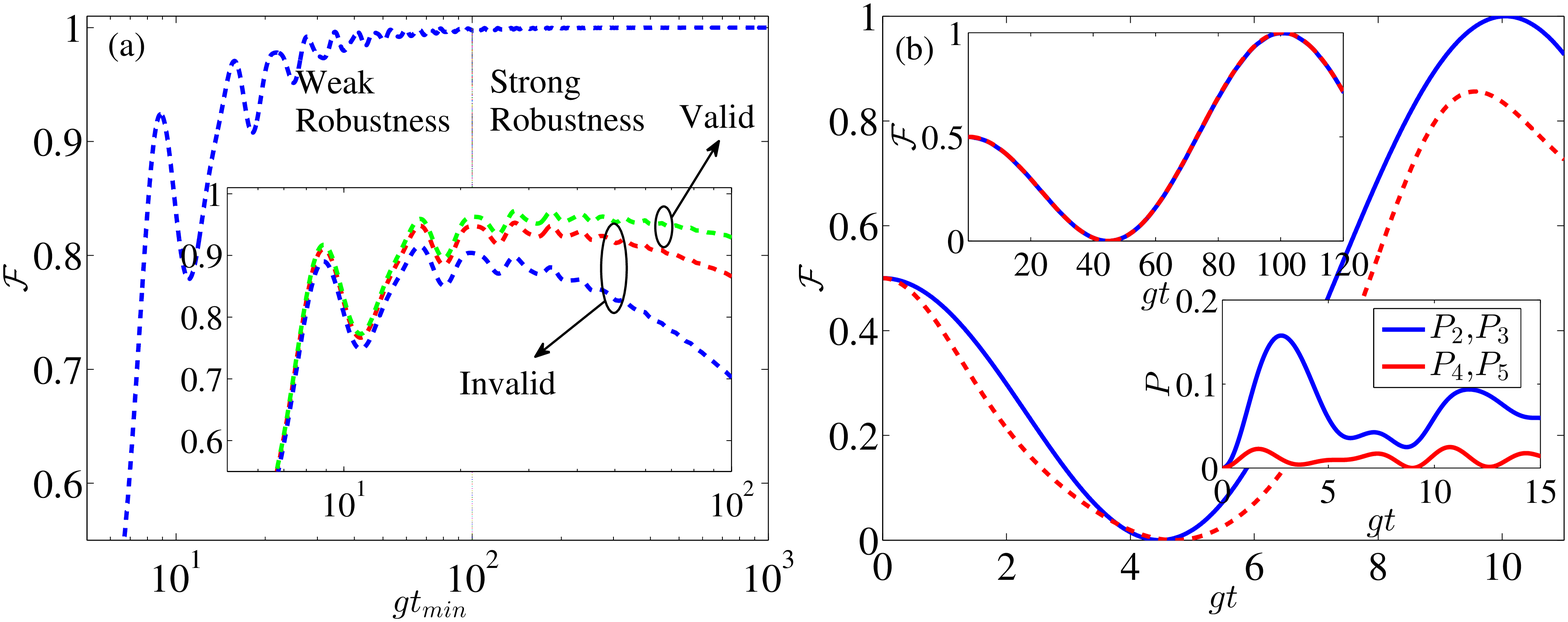}  
\end{minipage}
\caption{(a): The maximum fidelity and the corresponding $t_{min}$ with varied $\Omega_2/g$. Main figure exhibits the change of closed system and the inset shows the cases corresponding to $\gamma=0.0005$ (green), $\gamma=0.001$ (red) and $\gamma=0.002$ (blue), respectively. (b): Fidelity with the varied time under $\Omega_2/g=0.5$ (main figure) and $\Omega_2/g=0.05$ (the top inset). The blue line denotes the approximate solution and the red line is the exact solution. Another inset shows the probabilities of different Zeno subspaces. In this calculation, we set $g=\lambda=1$ for convenience.
\label{fig:fig2}}
\end{figure}

If $\Omega_{1,2}\ll g,\lambda$, the Hamiltonian can be expanded to the following complete set:
\begin{equation}
\begin{split}
\{&\vert\phi_1\rangle=\vert fg000\rangle,\,\vert\phi_2\rangle=\vert fe000\rangle,\,\vert\phi_3\rangle=\vert ff010\rangle\\
&\vert\phi_4\rangle=\vert ff100\rangle,\,\vert\phi_5\rangle=\vert ff001\rangle,\,\vert\phi_6\rangle=\vert gf000\rangle\\
&\vert\phi_7\rangle=\vert ef000\rangle\},
\label{eq:Basic}
\end{split}
\end{equation}
and one can diagonalize conveniently after neglecting $H_{laser}$. {What needs to be explained is that $\vert fg000\rangle$ denotes $\vert fg\rangle_{AB}\vert 00\rangle_{c_1c_2}\vert 0\rangle_{f}$ and other vectors in Eq. \eqref{eq:Basic} obey the same order.} Under these conditions, the whole Hilbert space can be divided into five Zeno subspaces \cite{24}: 
\begin{equation}
\begin{split}
&Z_1=\{\vert\psi_{1}\rangle,\vert\psi_{2}\rangle,\vert\psi_{3}\rangle\}\\
&Z_2=\{\vert\psi_{4}\rangle\}\\
&Z_3=\{\vert\psi_{5}\rangle\}\\
&Z_4=\{\vert\psi_{6}\rangle\}\\
&Z_5=\{\vert\psi_{7}\rangle\},
\label{eq:Zeno}
\end{split}
\end{equation}
corresponding to the eigenvalues $\zeta_{1,2,3}=0$, $\zeta_2=g$, $\zeta_3=-g$, $\zeta_4=\sqrt{g^2+2\lambda^2}$ and $\zeta_5=-\sqrt{g^2+2\lambda^2}$, respectively. Using the related derivations of Zeno dynamics (Eqs. (\ref{eq:subtime},\ref{eq:time})), the Hamiltonian in Eq. (\ref{eq:Hamilton}) can be rewritten in the form of: $H_{eff}=\Omega_2\delta\vert\psi_{1}\rangle\langle\psi_{2}\vert+\Omega_1\delta\vert\psi_{2}\rangle\langle\psi_{3}\vert+H.c.+\sum_{k=4}^7\zeta_k\vert\psi_k\rangle\langle\psi_k\vert$ \cite{25}, and it can be reduced to $H_{eff}=\Omega_2\delta\vert\psi_{1}\rangle\langle\psi_{2}\vert+\Omega_1\delta\vert\psi_{2}\rangle\langle\psi_{3}\vert+H.c.$ if the initial state is limited in the first Zeno subspace $Z_1$. $H_{eff}$ is just a $3\times 3$ matrix, therefore the evolution of the system state can be easily calculated and expressed as 
\begin{equation}
\begin{split}
\vert\psi(t)\rangle=&\dfrac{1}{\Omega^2}[(\Omega^2_1+\Omega^2_2\cos\Omega\delta t)\vert\psi_{1}\rangle-i\Omega\Omega_2\sin\Omega\delta t\vert\psi_{2}\rangle\\&+\Omega_1\Omega_2(-1+\cos\Omega\delta t)\vert\psi_{3}\rangle],
\label{eq:State}
\end{split}
\end{equation}
corresponding to the initial state $\vert\psi(0)\rangle=\vert fg\rangle_{AB}\vert00\rangle_{C_1C_2}\vert 0\rangle_f$. While the parameters are taken as $t_s=(2n+1)\pi/\Omega\delta$ and $\Omega_1=(\sqrt{2}-1)\Omega_2$, Eq. (\ref{eq:State}) becomes: $\vert\psi(t_s)\rangle=(\vert\psi_1\rangle+\vert\psi_3\rangle)/\sqrt{2}$ and the corresponding atoms are in the Bell state: $\vert\psi^+\rangle=(\vert fg\rangle+\vert gf\rangle)/\sqrt{2}$; On the contrary, if we set $t_s=(2n+1)\pi/\Omega\delta$ and $\Omega_1=(\sqrt{2}+1)\Omega_2$, the atoms are in the Bell state: $\vert\psi^-\rangle=(\vert fg\rangle-\vert gf\rangle)/\sqrt{2}$. 

Through analyzing above entanglement preparation scheme, we find that the fastest time for the initial state to achieve the maximum entangled state is $gt_{min}=\pi/\Omega\delta$. However, the Zeno dynamic requires that $\Omega_{1,2}\ll g,\lambda$, which will lead to too long $gt_{min}$. To explain this, we consider the first case and plot the fidelity $\mathcal{F}(\vert\psi^+\rangle,\rho)=\langle\psi^+\vert\rho\vert\psi^+\rangle$ \cite{26,27,FF} and the corresponding $gt_{min}$ with the varied $\Omega_2/g$ in Fig. \ref{fig:fig2}. In Fig. \ref{fig:fig2}(a), we show that the oscillating fidelity is lower than $35\%(85\%)$ when $gt_{min}<5(15)$. {Similarly to previous works, Fig.\ref{fig:fig2} shows that the fidelity does not decrease monotonically with increasing $\Omega$ when the approximate condition is destroyed. In time domain, this phenomenon is reflected in the oscillating fidelity when $gt_{min}$ is small.  The fidelity with intensive oscillation indicates that a tiny parameter deviation will affect the fidelity significantly, and the system actually corresponds to a state of weak robustness. On the contrary, the fidelity in a practical entanglement preparation scheme is required to be stable and to tend to the approximate solution (i.e., exhibits a strong robustness). And we further find that this requirement will not be satisfied until $gt=100$ \cite{r1}.} This long evolution time may make the scheme ineffective when the system interacts with the environment. In the inset in Fig. \ref{fig:fig2}(a), we also show system  evolution under the following non--Hermitian Hamiltonian
\begin{equation}
H'_{eff}=H_{eff}-i\sum\dfrac{\gamma_{j}}{2}\vert\phi_j\rangle\langle\phi_j\vert,
\label{eq:Hde}
\end{equation}
where $\gamma_{j}$ are the dissipation coefficients of atoms or optical fields. It illustrates that the entanglement preparation scheme is valid only if $\gamma/g\leqslant 0.0005$, which corresponds to a very demanding experimental condition. In Fig. \ref{fig:fig2}(b), we plot the $\mathcal{F}(t)$ under different $ \Omega_2/g$ and find that the system state is not limited within the subspace $Z_1$ but jumps into other Zeno subspaces if the Zeno conditions are destroyed. 
\begin{figure}[]
\centering
\begin{minipage}[b]{0.5\textwidth}  
\includegraphics[width=3.5in]{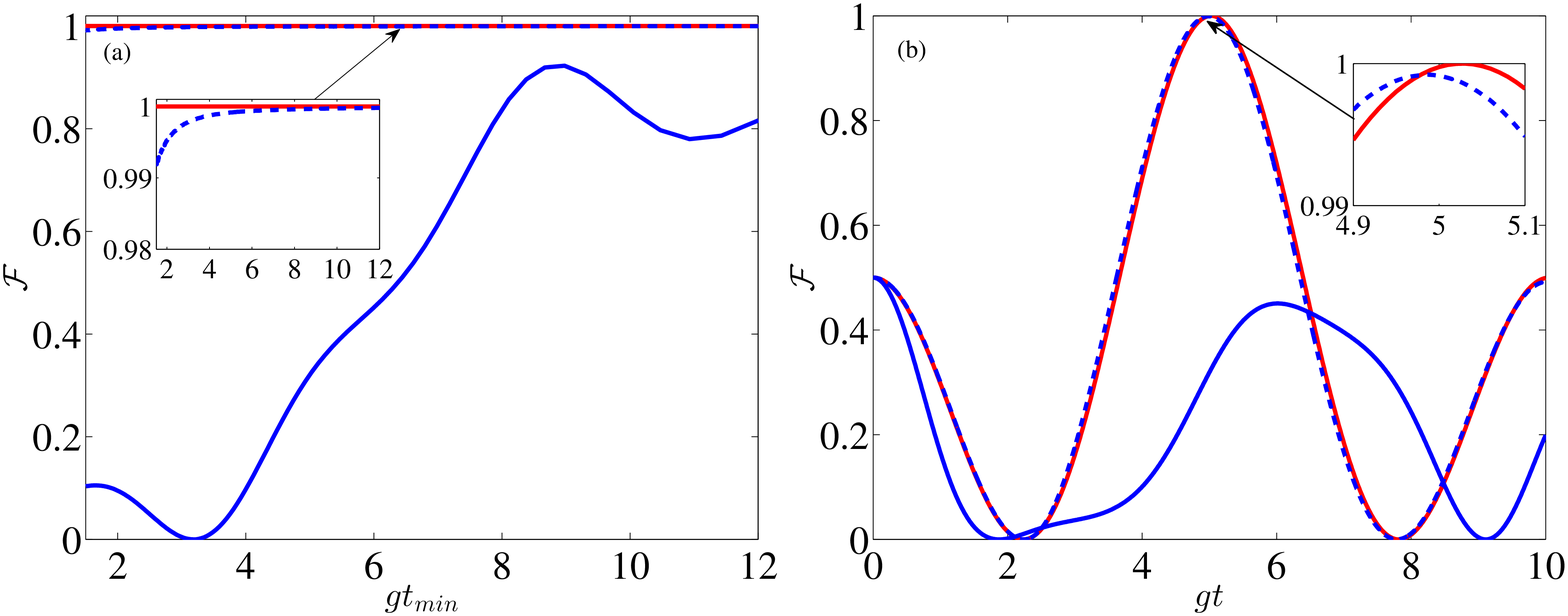}  
\end{minipage}
\caption{(a): The maximum fidelity and the corresponding $t_{min}$ with varied $\Omega_2/g$. (b): Fidelity evolution with varied $t$ under $\Omega_2/g=1$. The red line denotes the fidelity corresponding to ``rough'' acceleration scheme and the blue dotted and solid lines respectively are flexible acceleration and non-acceleration. Here we set $k_j=10$ and other parameters are the same as those in Fig. \ref{fig:fig2}.
\label{fig:fig3}}
\end{figure}  
  
As we discussed in Sec. \ref{subsection:General formalism}, the acceleration Hamiltonians of this scheme are respectively $H_{R}=UH_{eff}U^{\dagger}-H_{laser}=\Omega_2(\delta^2-1)\vert\phi_1\rangle\langle\phi_2\vert-\Omega_2\delta^2g/\lambda\vert\phi_1\rangle\langle\phi_5\vert+\Omega_2\delta^2\vert\phi_1\rangle\langle\phi_7\vert+\Omega_1\delta^2\vert\phi_2\rangle\langle\phi_6\vert-\Omega_1\delta^2g/\lambda\vert\phi_5\rangle\langle\phi_6\vert+\Omega_1(\delta^2-1)\vert\phi_6\rangle\langle\phi_7\vert+H.c.$ which corresponds to the ``rough'' acceleration, and $H_{F}=\sum_{j=0}^{n}u_{j}(t)H_{cj}$ which corresponds to the flexible acceleration. We firstly consider a set of complete $\{H_{cj}\}$ in order to compare two kinds of acceleration schemes, i.e., $H_{cj}$ are selected as $H_{c0}=H_{c1}=\vert\psi_1\rangle\langle\psi_4\vert+H.c.$; $H_{c(2,3,4)}=\vert\psi_1\rangle\langle\psi_{5,6,7}\vert+H.c.$; $H_{c(5,6,7,8)}=\vert\psi_2\rangle\langle\psi_{4,5,6,7}\vert+H.c.$; $H_{c(9,10,11,12)}=\vert\psi_3\rangle\langle\psi_{4,5,6,7}\vert+H.c.$, respectively, because complete $\{H_{cj}\}$ can ensure that the system is controlled to a maximum level without loss of fidelity.

In Fig. \ref{fig:fig3}, we show the contrast results corresponding to non-acceleration, ``rough'' acceleration and flexible acceleration, respectively. Fig. \ref{fig:fig3}(a) illustrates that the ``rough'' acceleration can hold $\mathcal{F}=1$ during the whole evolution period, contrarily, the fidelity of flexible acceleration keeps on increasing from $0.993$ to $1$. This phenomenon is natural because $H_R$ adds all of the approximated terms into the system whereas the Lyapunov control theory only provides $\mathcal{F}\rightarrow 1$ in limited time. However, the fidelity is still 
significantly greater than the one without any acceleration. In particular, in the range of $gt\in[1,5]$, the flexible acceleration can ensure that the fidelity is always greater than $99.9\%$ although the fidelity corresponding to non-acceleration is only $34.5\%$. In Fig. \ref{fig:fig3}(b), we show the time evolution of fidelity under $\Omega_2/g=1$. It can be directly observed that the fidelity of flexible acceleration exhibits a similar evolution with that of the ``rough" acceleration and the fidelity distortion is only $0.08\%$. 
\begin{figure}[]
\centering
\begin{minipage}[b]{0.5\textwidth}  
\includegraphics[width=3.5in]{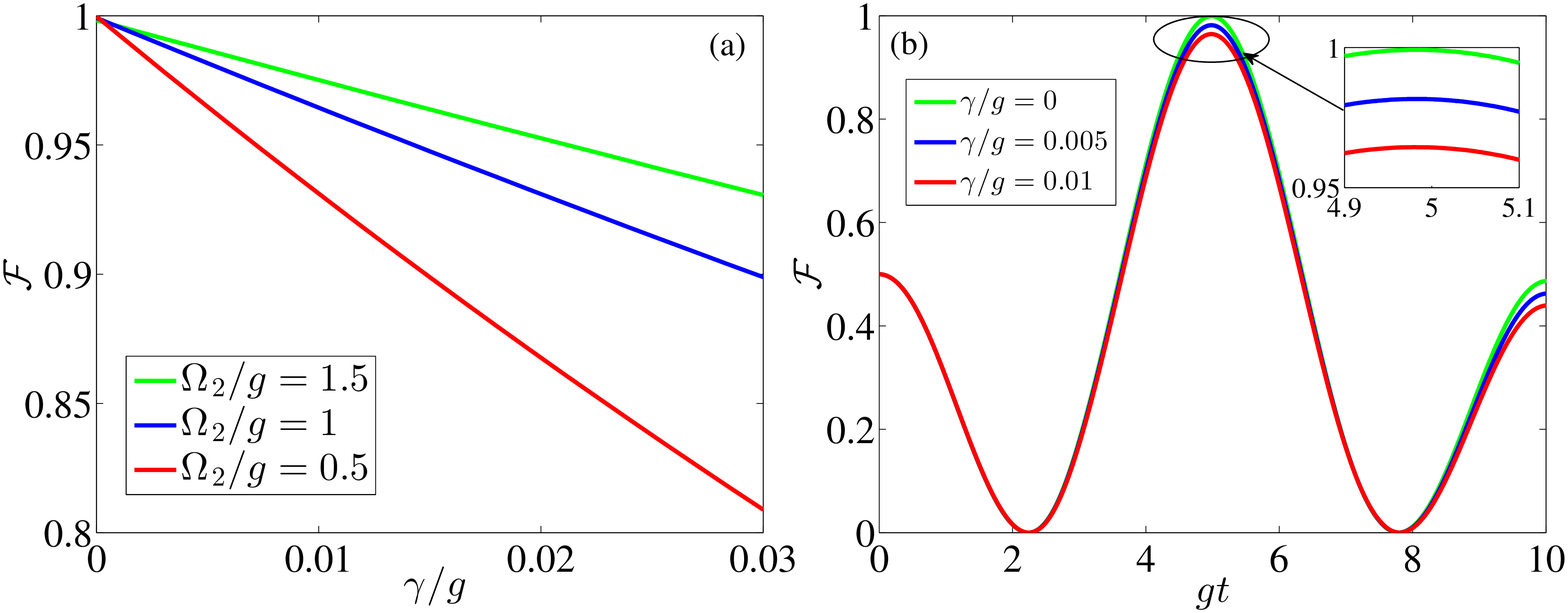}  
\end{minipage}
\caption{(a): The maximum fidelities with varied $\gamma/g$ under different $\Omega_2/g$. (b): Fidelity evolutions with varied $t$ under $\gamma/g$. All parameters in this simulation are the same as those in Fig. \ref{fig:fig3}.
\label{fig:fig4}}
\end{figure}
In Fig. \ref{fig:fig4}, we consider the influence of the environment interaction and the result shows that while $\Omega_2=g$, $\mathcal{F}$ will remain at the level of $90\%$ even though $\gamma/g=0.03$. This boundary is enlarged about $60$ times than that using the non-acceleration scheme, which results in easier implementation for our scheme in experiments.  \cite{17,29,30,31}.

For two $\Lambda$-type atoms in a real experiment, however, it can be known that both $H_R$ and $H_F$ do not exist because there are some non--physical interaction terms in their expressions (e.g., $\vert\phi_1\rangle\langle\phi_5\vert+H.c.,\vert\phi_1\rangle\langle\phi_7\vert+H.c.$, and so on). ``Rough'' acceleration can not fix this defect directly since $H_R$ is already an invariant function in a certain scheme. Contrarily, $H_F$ can be easily adjusted by selecting realizable $\{H_{cj}\}$ afresh. In general, a realizable $\{H_{cj}\}$ is usually an incomplete set of Hamiltonian. In the system shown  in Fig. \ref{fig:fig1}, for example, the $\{H_{cj}\}$ set constituted only by realizable interactions is:
\begin{equation}
\begin{split}
H_{cj}\in\{&H_{c1}=\vert\phi_1\rangle\langle\phi_2\vert+H.c.,H_{c2}=\vert\phi_6\rangle\langle\phi_7\vert+H.c.,\\
&H_{c3}=\vert\phi_2\rangle\langle\phi_3\vert+H.c.,H_{c4}=\vert\phi_4\rangle\langle\phi_7\vert+H.c.,\\
&H_{c5}=\vert\phi_3\rangle\langle\phi_5\vert+\vert\phi_4\rangle\langle\phi_5\vert+H.c.\}.
\label{eq:hrhr}
\end{split}
\end{equation}
{In this set, $H_{c1}\sim H_{c4}$ correspond to field--atom couplings and they can be achieved by adjusting the Rabi frequency and the cavity detuning of each transition processing. Correspondingly, $H_{c5}$ is cavity--fiber interaction and it can also be adjusted in some ring cavity systems. But this adjustment is uncommon, therefore we select $u_5$ as a constant for a general discussion.} This incomplete Hamiltonian can not contain all control paths between different Zeno subspaces, hence some fidelity distortions may exist here. Even so, the advantage of this scheme is obvious because all terms in Eq. (\ref{eq:hrhr}) can be realized in an experiment.
\begin{figure}[]
\centering
\begin{minipage}[b]{0.5\textwidth}  
\includegraphics[width=3.5in]{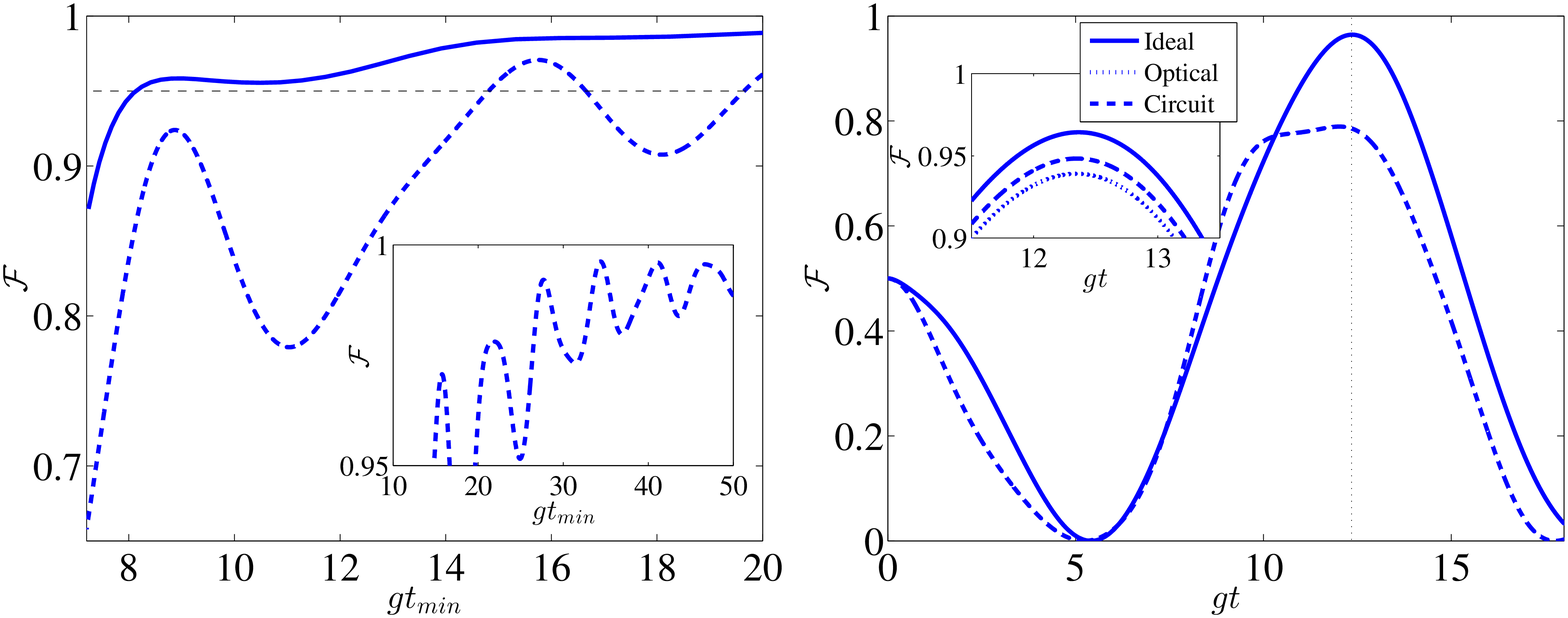}  
\end{minipage}
\caption{(a): The maximum fidelities and the corresponding $t_{min}$ with varied $\Omega_2/g$. (b): Fidelity evolutions with varied $t$ under the condition $t_{min}=12.4$ (main figure) and their performances under different experiment parameters (inset). The solid and dotted lines denote the fidelity corresponding to acceleration and non-acceleration. Here we set $H_{c0}=0$ and $k=0.6$, and other parameters in this simulation are the same as those in Fig. \ref{fig:fig3}.
\label{fig:fig5}}
\end{figure}
 
In Fig. \ref{fig:fig5}(a), we show that the fidelities of flexible acceleration can not fast and perfectly approach to $1$ such as that in Fig. \ref{fig:fig3}. {If we define such a standard, that is, the fidelity not only is greater than $95\%$ but also always keeps the status, i.e., system with stronger robustness}, the minimum evolution time corresponding to flexible acceleration is $gt_{min}=8.1$, and obviously, it is still nearly $60\%$ compression compared with $gt_{min}=20$ in non-acceleration scheme. We also plot fidelity evolutions in Fig. \ref{fig:fig5}(b) to show a significant promotion at $gt_{min}=12.4$ in this acceleration process. Here we also present a brief discussion about the actual effectiveness of our acceleration under following experiment parameters. Recent experiments of cavity QED system have achieved $(\kappa,\beta_c,\beta_f)/g=(0.0035,0.0047,0.0002)$ in Fabry--P\'erot cavity \cite{33,34}, and $(\kappa,\beta_c,\beta_f)/g=(0.0021,0.0004,0.0004)$ in circuit QED system\cite{35,36}. In Fig. \ref{fig:fig5}(b), we also illustrate that $\mathcal{F}\geqslant 93\%$ is still satisfied in those experiment systems even if $\Omega_2/g\in [0.4,0.6]$. Therefore, we believe our acceleration is feasible under present available experiment technology.

{Finally, we will analyze the realizability of the functions $u_j(t)$ in experiments. It should be stressed that the experimentally feasible $u_j(t)$ should at least meet the following two requirements. One is that the interaction corresponding to each $u_j(t)$ should be achievable and can be adjusted; Another is that all $u_j(t)$ should be of smooth waveforms, and it is better that $u_j(t)$ are constituted by some common waveforms without high-frequency oscillation (sine function, square pulses and Gaussian function for examples) \cite{6}. In our scheme, the only used time-dependent control field are $u_{1,2,3,4}(t)$ and it already has been discussed that those corresponding interactions should be explicit time--dependent. Therefore, $u_j(t)$ exactly satisfy the first requirement in our model.
\begin{figure}[]
\centering
\begin{minipage}[b]{0.5\textwidth}  
\includegraphics[width=3.5in]{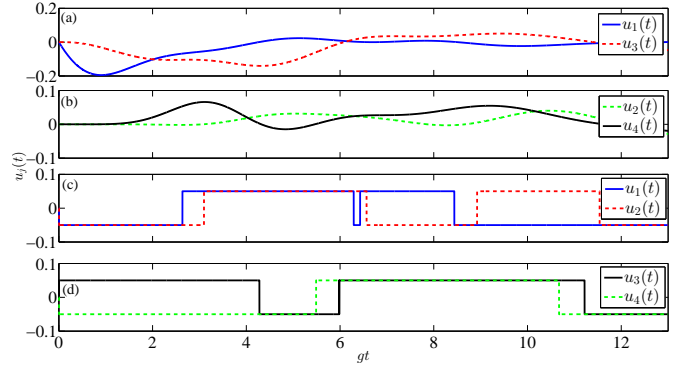}  
\end{minipage}
\caption{Waveforms of $u_j(t)$. (a) and (b) are initial designs corresponding to Eq. (\ref{eq:CON function1}); (c) and (d) are square pulses corresponding to Eq. (\ref{eq:CON functionsq}). Here all parameters in this simulation are the same as those in Fig. \ref{fig:fig3}.
\label{fig:fig6}}
\end{figure}
Considering the second requirement, we plot time evolutions of $u_j(t)$ in Fig. \ref{fig:fig6} (a) and (b) to show that $u_j(t)$ are smooth enough for the intensity adjustment \cite{6,37,38,39}. We also want to point out that $u_j(t)$ can be of various forms without being limited to an exclusive form since the Lyapunov control just needs to determine the positive or negative value instead of obtaining the concrete accurate value of $\dot{V}$. For example, we can intuitively set $u_{1,2,3,4}(t)$ as 
\begin{equation}
u_{j}=
\left\lbrace
\begin{array}{ll}
K\,\,\,\,\,\,\,\,\,\,\, if\,\,\,\,{\Trr(-i\rho[H_I^2,H_{cj}])}<0\\
\\
-K\,\,\,\,\,\,\, if\,\,\,\,{\Trr(-i\rho[H_I^2,H_{cj}])}>0,
\label{eq:CON functionsq}
\end{array}
\right.
\end{equation}
and consequently, $u_j(t)$ will be simplified as the square pulses without high frequency oscillation, which can be more easily implemented \cite{3,6,40} by just controlling the on/off of control fields with constant intensities. With these square pulses, the fidelities are still greater than $96.7{\%} $ at the $t_{min}=10.8$, which means not only $H_{cj}$ but also $u_j(t)$ can be selected flexibly. This flexibility ensures that our scheme can always be implemented experimentally.}
\section{Discussion and Outlook}
In this letter, we have proposed a flexible and realizable acceleration scheme to speed up Zeno dynamic passage that has already been used widely in QIP. Unlike the efforts to eliminate the effects of neglected terms, the basic idea of our acceleration is to drive the system for evolving within an appointed subspace. The acceleration Hamiltonian is discussed with a general form $H_F=\sum u_{j}(t)H_{cj}$ instead of the traditional fixed form $H_{R}=UH_{eff}U^{\dagger}-H$.  Thus, our acceleration Hamiltonian can be designed limberly by selecting different $H_{cj}$. Especially, one can always find such $\{H_{cj}\}$ set in which each element is reasonable and can be realized in experiments.  Our acceleration scheme has been applied on an entanglement preparation process, and the result shows that the flexible acceleration can shorten the evolution time $gt_{min}$ from $20$ to $8.1$ under the condition $\mathcal{F}\geqslant 95\%$. On the other hand, the requisite time for a high robustness is also reduced to $gt_{min}=12$, which is clearly shorter than the time $gt_{min}=100$ when there does not exist acceleration scheme. In addition, the restrictions of the decay rates are also relaxed by the acceleration process. It can be found that our flexible acceleration will provide a more feasible scheme if the acceleration field $u_{j}(t)$ are taken as Gaussian distributions. The possibility of the idea will be further verified in some subsequent researches.

\section*{Acknowledgement}
All authors thank Jiong Cheng, Wenzhao Zhang and Yang Zhang for the useful discussion. This research was supported by the National Natural Science Foundation of China (Grant No 11175033, No 11574041, No 11505024 and No 11447135.) and the Fundamental Research Funds for the Central Universities (DUT13LK05).


\begin{thebibliography}{99}
\bibitem{r1} {For cavity QED system, $g$ often is of the scale of MHz, and correspondingly, $t\sim 10^{-4}s$.}
\bibitem{24} In Eq. (\ref{eq:Zeno}), the concrete forms of the eigenvectors are:\\$\vert\psi_1\rangle=\vert\phi_1\rangle$\\$\vert\psi_2\rangle=\delta\left(\vert\phi_2\rangle-\dfrac{g}{\lambda}\vert\phi_5\rangle+\vert\phi_7\rangle\right)$\\$\vert\psi_3\rangle=\vert\phi_6\rangle$\\$\vert\psi_4\rangle=\dfrac{1}{2}(\vert\phi_2\rangle+\vert\phi_3\rangle-\vert\phi_4\rangle-\vert\phi_7\rangle)$\\$\vert\psi_5\rangle=\dfrac{1}{2}(\vert\phi_2\rangle-\vert\phi_3\rangle+\vert\phi_4\rangle-\vert\phi_7\rangle)$\\$\vert\psi_6\rangle=\delta\left(\dfrac{g}{2\lambda}\vert\phi_2\rangle+\dfrac{1}{2\delta}\vert\phi_3\rangle+\dfrac{1}{2\delta}\vert\phi_4\rangle+\vert\phi_5\rangle+\dfrac{g}{2\lambda}\vert\phi_7\rangle\right)$\\$\vert\psi_7\rangle=\delta\left(\dfrac{g}{2\lambda}\vert\phi_2\rangle-\dfrac{1}{2\delta}\vert\phi_3\rangle-\dfrac{1}{2\delta}\vert\phi_4\rangle+\vert\phi_5\rangle+\dfrac{g}{2\lambda}\vert\phi_7\rangle\right)$
\bibitem{25} For convenience, we set $\delta=\lambda/\sqrt{g^2+2\lambda^2}$ and $\Omega^2=\Omega_1^2+\Omega_2^2$ in following calculations.
\bibitem{28} $\hat{\gamma}=[0,2\kappa,\beta_c,\beta_c,4\beta_f,0,2\kappa]$, where $\kappa$ is the atomic spontaneous decay rate. $\beta_c$ and $\beta_f$ are the decay rates of the cavity and the fiber. For convenience, we set $\hat{\gamma}=\gamma[0,2,1,1,1,0,2]$.
\end{thebibliography}
\end{document}